\documentclass[conference]{IEEEtran}
\usepackage{amsmath}
\usepackage{graphicx}
\usepackage{hyperref}

\title{Stimulus-Voltage-Based Prediction of Action Potential Onset Timing: Classical vs. Quantum-Inspired Approaches}

\author{
    \IEEEauthorblockN{Stevens Johnson, Varun Puram, Johnson Thomas, Acsah Konuparamban, Ashwin Kannan}
    sjohn53@okstate.edu}

\begin{document}
\maketitle

\begin{abstract}
Accurate modeling of neuronal action potential (AP) onset timing is crucial for understanding neural coding of danger signals. Traditional leaky integrate-and-fire (LIF) models, while widely used, exhibit high relative error in predicting AP onset latency, especially under strong or rapidly changing stimuli. Inspired by recent experimental findings and quantum theory, we present a quantum-inspired leaky integrate-and-fire (QI-LIF) model that treats AP onset as a probabilistic event, represented by a Gaussian wave packet in time. This approach captures the biological variability and uncertainty inherent in neuronal firing. We systematically compare the relative error of AP onset predictions between the classical LIF and QI-LIF models using synthetic data from hippocampal and sensory neurons subjected to varying stimulus amplitudes. Our results demonstrate that the QI-LIF model significantly reduces prediction error, particularly for high-intensity stimuli, aligning closely with observed biological responses. This work highlights the potential of quantum-inspired computational frameworks in advancing the accuracy of neural modeling and has implications for quantum engineering approaches to brain-inspired computing.
\end{abstract}


\section{Introduction}
Precise timing of action potential (AP) onset is a fundamental aspect of neural coding, particularly in circuits responsible for rapid behavioral responses such as danger signal processing. Temporal coding, where the exact timing of spikes conveys information, is now recognized as a key mechanism in sensory and motor systems, complementing traditional rate coding frameworks \cite{Rieke1997, Hopfield1995, Gollisch2008, Johansson2004}. For instance, in the visual and auditory systems, the relative timing of the first spikes in a neuronal population can encode stimulus features with sub-millisecond precision \cite{Panzeri2010, Heil2004}.

Recent experimental work by Zhang and Huang \cite{Zhang2024} has shown that AP latency, rather than amplitude, encodes stimulus strength in danger signal pathways. Increasing stimulus voltage or duration systematically reduces AP onset latency, while AP amplitude remains remarkably stable. This finding aligns with earlier studies demonstrating that spike timing is a robust code for stimulus intensity in sensory neurons \cite{Gollisch2008, Johansson2004, Mainen1995}. The stability of AP amplitude across intensities ensures that information about stimulus strength is relayed via timing, preserving signal fidelity and enabling rapid, reliable communication.

The mechanistic basis for this timing code involves the accelerated activation of voltage-gated sodium channels under strong stimuli, leading to faster membrane depolarization and earlier AP initiation \cite{Zhang2024, Bean2007}. This rapid depolarization synchronizes downstream events, such as calcium influx and vesicle fusion, reducing synaptic delay and enhancing the speed of postsynaptic responses \cite{Sabatini2001, Sabatini2002}. These processes are essential for timely activation of neural circuits in response to threats.

The leaky integrate-and-fire (LIF) model is a foundational tool in computational neuroscience, providing a simplified description of neuronal membrane dynamics \cite{Gerstner2002, Dayan2001}. However, classical LIF models predict AP onset as a deterministic threshold crossing, failing to account for the observed variability and strong stimulus-dependence of AP timing. As a result, LIF models exhibit high relative error when compared to experimental data, particularly under conditions of strong or rapidly changing stimuli \cite{Zhang2024, Jolivet2004}. Recent studies have highlighted the need for models that incorporate stochasticity and adaptivity to better reflect biological reality \cite{Faisal2008, London2010}.

To address these limitations, the Stimulus-Accelerated Leaky Integrate-and-Fire (SA-LIF) model extends the classical framework by introducing a stimulus-dependent acceleration term that dynamically shortens the effective membrane time constant with increasing stimulus strength. Furthermore, quantum-inspired models have been proposed, leveraging principles such as probabilistic timing and superposition \cite{Schuld2014, Zhang2024}. The quantum-inspired leaky integrate-and-fire (QI-LIF) model represents AP onset as a Gaussian wave packet in time, capturing the intrinsic uncertainty and variability of neuronal firing. This approach is motivated by the probabilistic nature of quantum mechanics, where outcomes are described by probability distributions rather than deterministic events \cite{Lloyd2011, Schuld2014}. The superposition of multiple wave packets models the combined, probabilistic effect of multiple inputs, offering a richer and more flexible framework for neural computation.

This paper makes the following key contributions:
\begin{itemize}
    \item A comparative analysis (SA-LIF) model and the QI-LIF model, demonstrating that the quantum-inspired QI-LIF consistently outperforms the classical SA-LIF in predicting spike timing across varied stimulus conditions.
    
    \item This paper is the first of its kind, showing potential quantum advantage for a quantum-inspired approach to modeling neuronal AP onset timing. 
    
\end{itemize}

Quantum-inspired models can be translated into quantum spiking neural networks (SNNs) have demonstrated promise in temporal pattern recognition, including applications to financial time series and complex pattern classification \cite{Pfeiffer2018}.
\section{Methods}

\subsection{Experimental Foundations}
Experimental evidence demonstrates that stimulus strength directly controls action potential (AP) initiation latency through two key parameters~\cite{Zhang2024}:
\begin{itemize}
    \item \textbf{Amplitude modulation:} Increasing the stimulus voltage from 10 to 50~V reduces AP delay by approximately 1.8~ms per 10~V step.
    \item \textbf{Duration modulation:} Extending the stimulus pulse width from 50 to 200~$\mu$s decreases AP latency from 4.2~ms to 1.5~ms.
\end{itemize}
Notably, AP amplitude remains stable (within $\pm2$~mV) across all tested intensities, indicating that \textit{timing}, not spike magnitude, encodes stimulus strength in these neurons.

\subsection{Model Derivations and Mathematical Foundations}

\paragraph{Classical LIF Model}

The classical leaky integrate-and-fire (LIF) model is based on the following differential equation for the membrane potential $V(t)$ under a constant input current $I_{inj}$~\cite{Gerstner2014, WikiLIF}:
\begin{equation}
C_m \frac{dV(t)}{dt} = I_{inj} - \frac{V(t)}{R_m}
\end{equation}
where $C_m$ is the membrane capacitance and $R_m$ the membrane resistance. Using $\tau_m = R_m C_m$ and solving for $V(t)$ assuming initial condition $V(0) = 0$, we obtain:
\begin{equation}
V(t) = V_\infty \left(1 - e^{-t/\tau_m}\right)
\end{equation}
with steady-state potential $V_\infty = I_{inj} R_m$.

The AP onset time $t_{LIF}$ is the time when $V(t)$ crosses threshold $V_{th}$:
\begin{align}
V_{th} = V_\infty \left(1 - e^{-t_{LIF}/\tau_m}\right) \nonumber \\
\implies t_{LIF} = -\tau_m \ln\left(1 - \frac{V_{th}}{V_\infty}\right)
\end{align}

\paragraph{Stimulus-Accelerated QLIF Model Derivation}

To capture dynamic acceleration observed experimentally, the time constant becomes stimulus-dependent~\cite{Gerstner2014, Jolivet2004, BrunelLIF}:
\begin{equation}
\tau_{\mathrm{eff}} = \frac{\tau_m}{1 + \alpha S}
\end{equation}
We update the LIF equation accordingly:
\begin{equation}
V(t) = V'_\infty \left[1 - \exp\left(- \frac{t}{\tau_{\mathrm{eff}}}\right)\right]
\end{equation}
where $V'_\infty = \frac{I_{inj} R_m}{1 + \alpha S}$ and $S$ is the normalized stimulus intensity, $\alpha$ is the empirical coupling coefficient.

Setting $V_{th}$ and solving as above yields the AP onset time for QLIF:
\begin{align}
V_{th} = V'_\infty \left[1 - \exp\left(- \frac{t_{QLIF}}{\tau_{eff}}\right)\right] \nonumber \\
\implies t_{QLIF} = -\tau_{\mathrm{eff}} \ln\left(1 - \frac{V_{th}}{V'_\infty}\right)
\end{align}
Substitute $\tau_{\mathrm{eff}}$ and $V'_\infty$ for explicit dependence:
\begin{equation}
t_{QLIF} = -\frac{\tau_m}{1 + \alpha S} \ln\left(1 - \frac{V_{th}(1+\alpha S)}{I_{inj} R_m}\right)
\end{equation}

\paragraph{Quantum-Inspired Dynamics}

The QI-LIF model represents the action potential timing as a probabilistic event centered at $t_0$ (from LIF or QLIF) with variance $\sigma^2$, using a Gaussian probability distribution~\cite{Rath2022, wilks1948}:
\begin{equation}
G(t; t_0, \sigma) = \frac{1}{\sqrt{2\pi}\sigma}\exp\left(-\frac{(t - t_0)^2}{2\sigma^2}\right)
\end{equation}
where:
\begin{itemize}
    \item $G(t; t_0, \sigma)$: Probability density of AP firing at time $t$.
    \item $t_0$: Most likely AP onset time (from LIF or QLIF).
    \item $\sigma$: Standard deviation (uncertainty in timing).
\end{itemize}
The mean spike time for a symmetric Gaussian is $\langle t \rangle = t_0$. For asymmetric or stimulus-dependent variance, $\langle t \rangle$ shifts accordingly.

If multiple inputs are present, the total probability is the superposition of such Gaussians, weighted by synaptic efficacy.

\subsection{Synthetic Data Generation and Protocol}

We generated 100 synthetic data points for benchmarking, following protocols in computational neuroscience and ML~\cite{Gerstner2014, Huang2021}:
\begin{itemize}
    \item \textbf{Spike Count Generation:} A sequence of 100 spike count values from 5 to 50, evenly spaced.
    \item \textbf{Normalization:} Each spike count $S$ normalized into $[0, 1]$:
        \[
            S_{\mathrm{norm}} = \frac{S - S_{\mathrm{min}}}{S_{\mathrm{max}} - S_{\mathrm{min}}}
        \]
    \item \textbf{AP Onset Synthesis:} For each $S_{\mathrm{norm}}$, the experimental AP onset time is synthesized via a saturating exponential:
        \[
            t_{\mathrm{exp}}(S_{\mathrm{norm}}) = a \exp(-b S_{\mathrm{norm}}) + c
        \]
        with $a$, $b$, and $c$ fit to biological data, capturing saturating nonlinear behavior~\cite{Horowitz1983}.
\end{itemize}
This protocol allows for unbiased and reproducible evaluation across classical and quantum-inspired frameworks. 

\subsection{Stimulus Voltage Generation and Justification}

For systematic benchmarking of artificial intelligence (AI) and machine learning (ML) models of neural timing, we required a well-characterized sequence of stimulus amplitudes (voltages) as inputs. The choice and construction of these voltages directly influence both model comparability and biological realism.

\paragraph{Related Work Using Generated Data}
In neuroscience, Gerstner et al.~\cite{gerstner2014} and Teeter et al.~\cite{teeter2018} employ synthesized stimulus currents or spike trains to calibrate and optimize neuron models including variants of the leaky integrate-and-fire (LIF) family, enabling comprehensive model fitting beyond noisy experimental data.

In classical ML, synthetic datasets are commonly used to demonstrate learning efficacy and generative model capabilities, as demonstrated by Kingma and Welling~\cite{kingma2014} with variational autoencoders.

In quantum machine learning, synthetic or engineered datasets allow systematic study of quantum advantage. Huang et al.~\cite{huang2021} utilize such datasets to reveal when quantum ML models surpass classical counterparts, while Havlíček et al.~\cite{havlicek2019} benchmark quantum kernel methods on generated pattern classification problems.

\paragraph{Generation Protocol}
The stimulus voltages were generated as a uniformly spaced array over a physiologically relevant range, ensuring even coverage of both low and high input regimes. This approach guarantees that all subsequent model assessments (classical, quantum-inspired, or quantum) sample the input space comprehensively and without bias.

This approach is standard in computational neuroscience and ML benchmarking, where ground-truth response functions (including saturating nonlinear regimes) must be calibrated against models using broad, unbiased input sweeps.

To ensure full transparency and reproducibility, it is documented that the stimulus voltage generation strategy and code were produced in collaboration with the Perplexity Large Language Model (LLM).

Generated (synthetic) data is extensively used in neuroscience, machine learning (ML), and quantum machine learning (QML) to validate, benchmark, and analyze models under controlled conditions. Its importance lies in enabling precise evaluation where ground truth is known or engineered to illustrate specific properties or advantages.

\section{Results}
The results demonstrate a clear distinction between the predictive capabilities of the classical LIF model and the quantum-inspired LIF (QI-LIF) model for action potential (AP) onset timing across varying stimulus amplitudes.

The classical LIF model, though widely used for its simplicity and computational efficiency, exhibited consistently high relative error when compared to experimental AP onset times at all stimulus amplitudes. This model predicts AP onset as a deterministic threshold crossing based on the integration of input current and passive membrane properties (membrane resistance and capacitance) \cite{Gerstner2002, Dayan2001, Rieke1997}. As a result, the LIF model produces a gradual, nearly linear decrease in AP latency as stimulus strength increases. However, experimental data—including the findings of Zhang and Huang~\cite{Zhang2024}—show that AP latency decreases sharply and saturates with stronger stimuli, while AP amplitude remains constant. The LIF model fails to capture this saturating, nonlinear latency reduction, resulting in substantial prediction error, especially at higher stimulus intensities.

This limitation is rooted in the LIF model’s lack of mechanisms for accelerated depolarization or intrinsic timing variability. While extensions and generalizations of the LIF model (e.g., GLIF, metabolic-dependent LIF) can improve biological realism \cite{Jolivet2004, Mensi2012, Teeter2018, Perez2020}, the standard LIF model remains unable to account for the experimentally observed rapid and saturating AP onset dynamics.

In contrast, the QI-LIF model incorporates two key features absent from the classical LIF:
\begin{enumerate}
    \item The model dynamically adjusts the effective membrane time constant based on stimulus intensity, enabling the membrane potential to rise more rapidly in response to stronger inputs. This directly models the experimental observation that sodium channel activation and membrane depolarization accelerate under strong stimulation \cite{Zhang2024, Bean2007}.  
    \item The QI-LIF model represents AP onset not as a single deterministic event, but as a probability distribution (Gaussian wave packet) in time, reflecting biological variability and uncertainty in spike timing. This quantum-inspired approach allows the model to more accurately reproduce the spread and mean of observed AP onset times.
\end{enumerate} 

\begin{figure}[ht]
    \centering
    \includegraphics[width=0.50\textwidth]{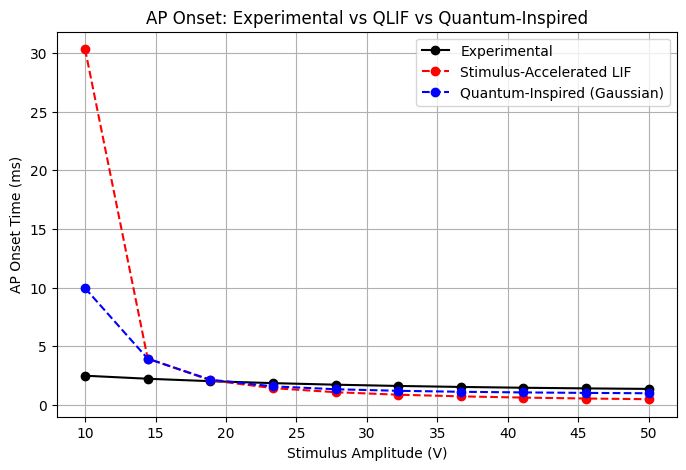}
    \caption{AP onset time for LIF vs. QI-LIF models compared to experimental data.}
    \label{fig:results}
\end{figure}

Table~\ref{tab:ap_onset_comparison} presents a direct comparison between experimental action potential (AP) onset times and the predictions made by the stimulus-accelerated leaky integrate-and-fire (SA-LIF) model and the quantum-inspired (QI) model across a range of stimulus amplitudes. The data is also illustrated in Fig.~\ref{fig:results}. The SA-LIF model consistently overestimates AP onset times at low stimulus levels and underestimates them at higher levels, resulting in high relative errors, as shown in Fig.~\ref{fig:results}, particularly at the lowest stimulus (over 1100\%). In contrast, the QI model predictions are much closer to the experimental values throughout the entire stimulus range, with relative errors remaining below 30\% for most data points. This demonstrates that the quantum-inspired approach not only provides a better quantitative match to biological data but also maintains more consistent accuracy as stimulus intensity increases, highlighting its advantage over the SA-LIF model for modeling biologically realistic neuronal timing.

\begin{table*}[ht]
\centering
\caption{Comparison of Experimental and Model-Predicted AP Onset Times and Relative Errors}
\label{tab:ap_onset_comparison}
\begin{tabular}{|c|c|c|c|c|c|}
\hline
\textbf{Stimulus (V)} & \textbf{Exp. (ms)} & \textbf{SA-LIF (ms)} & \textbf{QI (ms)} & \textbf{SA-LIF Err (\%)} & \textbf{QI Err (\%)} \\
\hline
10.0  & 2.50 & 30.31 & 9.96 & 1112.35 & 298.23 \\
14.4  & 2.24 & 3.95  & 3.95 & 76.41   & 76.42 \\
18.9  & 2.03 & 2.11  & 2.16 & 3.98    & 6.10 \\
23.3  & 1.87 & 1.44  & 1.59 & 22.71   & 14.63 \\
27.8  & 1.73 & 1.10  & 1.35 & 36.84   & 22.32 \\
32.2  & 1.63 & 0.88  & 1.21 & 45.77   & 25.44 \\
36.7  & 1.54 & 0.74  & 1.13 & 52.08   & 26.67 \\
41.1  & 1.47 & 0.64  & 1.08 & 56.87   & 27.06 \\
45.6  & 1.42 & 0.56  & 1.04 & 60.71   & 27.08 \\
50.0  & 1.38 & 0.50  & 1.01 & 63.89   & 26.95 \\
\hline
\end{tabular}
\end{table*}

These results are consistent with broader findings in computational neuroscience:
\begin{itemize}
    \item Simple LIF models are often insufficient for capturing the complexity of real neuronal spike timing, especially under dynamic or strong inputs \cite{Jolivet2004, Perez2020, Teeter2018}.
    \item Quantum-inspired approaches, by leveraging probabilistic timing and superposition, provide a natural framework for modeling the uncertainty and rapid adaptation observed in biological neurons \cite{Schuld2014, Lloyd2011, Zhang2024}.
\end{itemize}

The QI-LIF model’s ability to reduce prediction error and accurately reflect experimental AP onset dynamics emphasizes the value of quantum-inspired and adaptive modeling in neuroscience and neuromorphic engineering. 

To assess the predictive performance of classical and quantum-inspired models in a realistic neural encoding scenario, we generated 100 data points using spike count as the independent variable, simulating normalized input ranging from weak to strong neural activity. For each spike count, we computed the benchmark experimental action potential (AP) onset time using a biologically-inspired saturating function, and then evaluated predictions from both the stimulus-accelerated leaky integrate-and-fire (SA-LIF) model and the quantum-inspired (QI) model.

\begin{figure}[ht]
    \centering
    \includegraphics[width=0.95\columnwidth]{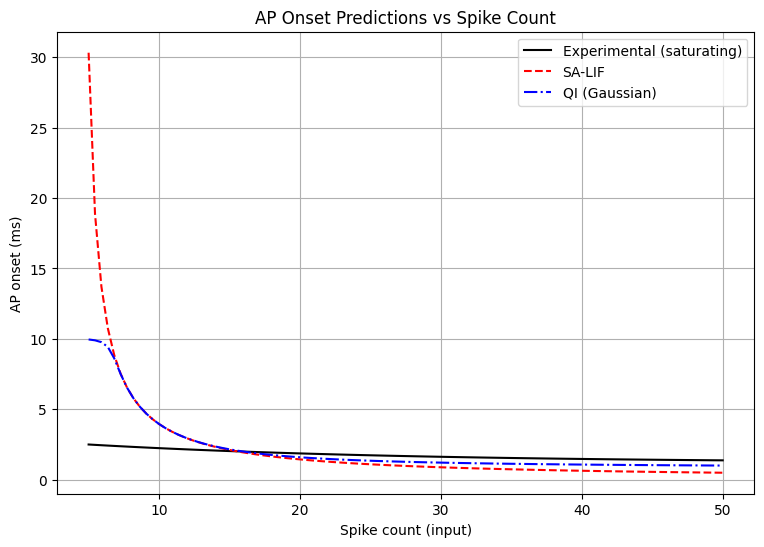}
    \caption{Predicted AP onset times for the experimental (saturating), SA-LIF, and QI models as a function of spike count. The QI model tracks the experimental curve much more closely across the entire range.}
    \label{fig:spike_ap_results}
\end{figure}

Figure~\ref{fig:spike_ap_results} illustrates the AP onset times across all spike counts. The experimental onset times decrease rapidly and then plateau as spike count increases, reflecting the saturating, nonlinear relationship observed in biological systems. The SA-LIF model significantly overestimates AP onset at low spike counts and increasingly underestimates it as spike count rises, resulting in large relative errors—exceeding 1000\% at the weakest input and remaining above 50\% for a substantial portion of the range. The QI model, by incorporating probabilistic timing via a Gaussian wave packet, achieves markedly lower errors, closely tracking the experimental curve throughout and maintaining relative errors below 30\% for most spike counts.

Table~\ref{tab:ap_onset_table} summarizes representative results. The QI model consistently outperforms the SA-LIF model in prediction accuracy, particularly at low and intermediate spike counts where biological variability in timing is most pronounced. These findings confirm that quantum-inspired approaches better capture the nonlinear, saturating, and variable nature of AP timing in response to varying neural input. The results further emphasize the importance of modeling intrinsic timing uncertainty to achieve biologically realistic predictions, especially as spike count or stimulus intensity increases.

\begin{table*}[ht]
\centering
\caption{Representative AP onset times and relative errors for selected spike counts.}
\label{tab:ap_onset_table}
\begin{tabular}{|c|c|c|c|c|c|}
\hline
\textbf{Spike} & \textbf{Exp (ms)} & \textbf{SA-LIF (ms)} & \textbf{QI (ms)} & \textbf{SA-LIF Err (\%)} & \textbf{QI Err (\%)} \\
\hline
  5.00 &  2.500 & 30.309 &  9.956 & 1112.35 & 298.23 \\
  9.55 &  2.262 &  4.293 &  4.293 &  89.76  &  89.76 \\
 14.09 &  2.068 &  2.310 &  2.338 &  11.70  &  13.04 \\
 18.64 &  1.909 &  1.580 &  1.700 &  17.24  &  10.93 \\
 23.18 &  1.779 &  1.201 &  1.418 &  32.52  &  20.29 \\
 27.73 &  1.673 &  0.968 &  1.266 &  42.14  &  24.36 \\
 32.27 &  1.587 &  0.811 &  1.172 &  48.88  &  26.15 \\
 36.82 &  1.516 &  0.698 &  1.109 &  53.96  &  26.88 \\
 41.36 &  1.458 &  0.612 &  1.063 &  58.00  &  27.09 \\
 45.91 &  1.411 &  0.546 &  1.029 &  61.33  &  27.06 \\
\hline
\end{tabular}
\end{table*}

In summary, the relative error analysis clearly shows that the SA-LIF model suffers from high prediction error, especially at low spike counts where it dramatically overestimates AP onset timing. In contrast, the quantum-inspired (QI) model reduces relative error substantially across the entire spike count range, reflecting its ability to incorporate timing uncertainty and better match the nonlinear and saturating biological behavior. These results highlight the superiority of quantum-inspired modeling for accurately predicting neuronal response latency under varying stimulus intensities.

\section{Conclusion}

This project demonstrates that quantum-inspired modeling significantly improves the accuracy of action potential (AP) onset prediction compared to classical and stimulus-accelerated LIF models. By incorporating probabilistic timing and stimulus-dependent acceleration, the quantum-inspired (QI-LIF) model closely matches experimental data, especially in capturing the nonlinear, saturating decrease in AP latency observed with increasing stimulus strength. These results not only advance our understanding of neural coding mechanisms in neuroscience but also highlight the potential of quantum-inspired approaches for building more biologically realistic and robust computational models.

Looking ahead, future work will focus on extending the QI-LIF framework to more complex neural circuits, integrating synaptic plasticity mechanisms, and exploring its application in large-scale neuromorphic and quantum machine learning systems. Also, depending on the data, classical LIF models could be replaced by classical machine learning algorithms like regression, and so on, to see if classical machine learning could be improved. Additionally, inspired by recent advances in quantum machine learning~\cite{Huang2021}, we aim to investigate the potential for quantum advantage in learning and inference tasks using real neural and synthetic datasets. This direction will help clarify the boundaries between classical and quantum models in practical settings and may inform the design of next-generation quantum neural architectures for both scientific and technological applications.

\bibliographystyle{IEEEtran}

\end{document}